\documentclass{elsart}
\usepackage{graphicx}

\begin{document}

\begin{frontmatter}

\title{Elliptic Flow of Rare High-Momentum Probes in Nuclear Collisions}

\author{W. Liu}
\address{Cyclotron Institute and Physics Department, Texas A$\&$M
University, College Station, Texas 77843-3366}
\author{R. J Fries}
\address{Cyclotron Institute and Physics Department, Texas A$\&$M
University, College Station, Texas 77843-3366}
\address{RIKEN/BNL Research Center, Brookhaven National Laboratory,
Upton, NY 11973}

\begin{abstract}
In high energy nuclear collisions the leading parton of a jet can change
flavor through interactions with the surrounding medium. This can considerably
boost the relative yield of rare high momentum particles, like strange
quarks and photons. We revisit these jet conversions and discuss
implications for the azimuthal asymmetry $v_2$ of rare probes. We predict
that the $v_2$ of kaons at RHIC falls significantly
below that of pions and protons at high transverse momentum. Experimental
information on the relative $v_2$ of strange hadrons could provide
constraints on the mean free path of fast quarks and gluons in a quark
gluon plasma.
\end{abstract}


\end{frontmatter}

\section{Rare High-Momentum Probes}

In collisions of nuclei at very high energies, the bulk of the matter is
expected to form a quark gluon plasma (QGP) in which partons are deconfined
and close to thermal equilibrium. Experiments at the Relativistic Heavy
Ion Collider (RHIC) and, soon, at the Large Hadron Collider (LHC) are
carried out to test the properties of quark gluon plasma \cite{whitepaper:05}.
Quantum chromodynamics (QCD) predicts that in some collisions initial
hard interactions of two partons take place which generate a final
state of partons with high transverse momentum $p_T$, which eventually
hadronize into energetic jets of hadrons. It had been realized early on
that these jets are an ideal probe for the medium formed in heavy ion
collisions \cite{Wang:1991xy,BDMPS:96,Zakharov:96,Wiedemann:2000tf,gyulassy,AMY:02,Wang:2003aw}.

The early years of experiments at RHIC have accumulated a large
sample of data which confirms the formation of a deconfined phase of
quarks and gluons. Measurements of high-$p_T$ hadrons played an important
role. A large suppression of high-$p_T$ hadrons and an almost complete
extinction of away-side jet correlations has been seen, emphasizing
the opacity of quark gluon plasma for fast partons.
\cite{whitepaper:05}.

Recently, it has been suggested that the study of energy loss and
suppression of high-$p_T$ particles should be augmented by a tracking of their
flavor to obtain additional, complementary information \cite{weiliu}.
Obviously, fast quarks and gluons traveling through nuclear matter can
change identity through conversion processes. This was first studied in
the context of conversions of light quarks into real and virtual photons
\cite{fries1,FMS:05,SGF:02,GAFS:04,simon,Turbide:2007mi} and the conversion of
of gluons into quarks and vice versa \cite{weiliu1,cmk,Schafer:2007xh}.
Note that to simplify notations our definition of flavor here refers to the
identity of any particle that can be produced at high $p_T$, including photons
and gluons. The most important consequences from these studies were worked
out in detail in Ref.\ \cite{weiliu}: (i) the concept of a
fixed flavor for the leading parton of a jet is ill-defined in a medium;
(ii) rare high-$p_T$ flavors like photons or strange quarks (the latter
only at RHIC energies) can be significantly enhanced through the coupling
to a chemically equilibrated medium; (iii) measurements of these rare probes
could provide information about the mean free path $\lambda$ of fast partons
in the medium.

In summary, measurements of excess photons and dileptons at intermediate and
high $p_T$ and changes to the hadron abundances in the jet fragmentation
region ($p_T >$ 6 GeV/$c$ at RHIC) could provide valuable information about
the quark gluon plasma formed in these collisions. In Ref.\
\cite{Sapeta:2007ad} it was pointed out that changes in hadron chemistry
at high $p_T$ could also come about through changed multiplicities in a
jet cone in nuclear collisions. This effect should
be distinguished from the mechanism discussed here which is based on flavor
changes of the leading parton. Possible effects for heavy quarks were
studied by us in \cite{Liu:2008bw}.

\section{Elliptic Flow}

In this Letter we want to discuss the azimuthal anisotropy $v_2$, often
referred to as elliptic flow, of rare high-$p_T$ probes. $v_2$ for a given
hadron species is defined as the second Fourier coefficient in the
decomposition of its transverse momentum spectra in terms of the
azimuthal angle $\phi$ with respect to the reaction plane
\begin{eqnarray}
 \frac{dN}{p_Tdp_Td\phi} = \frac{dN}{2\pi p_Tdp_T}\left[1+ 2v_2(p_T)\cos
  (2\phi)+ \mathcal{O}(4\phi)\right] \, .
\end{eqnarray}
It has been first pointed out in \cite{Turbide:2005bz} that real and
virtual photons produced from conversions of quark and gluon jets exhibit
a $v_2$ which has a negative sign, unlike all other particle production
mechanisms discussed up to that point. This contribution to the total photon
or dilepton spectrum has to be folded together with other photon sources which
has been the subject of an increasing number of studies very recently
\cite{Turbide:2007mi,Chatterjee:2005de,Kopeliovich:2007sd}.

We want to generalize the original argument in \cite{Turbide:2005bz}.
We point out that for any high-$p_T$ probe $R$ which
receives an additional contribution $\Delta R > 0$ through interactions
of other probes $J$ with the medium $M$, the additional yield $\Delta R$
exhibits an azimuthal asymmetry which is offset by $\pi/2$ with respect
to the reaction plane, i.e. has $v_2^{\Delta R} < 0$. Whether this negative
elliptic flow is visible in experiment depends on the probe. Typically, $v_2$
is positive for other sources of $R$, outshining the contribution from jet
conversions. But if $R$ is a rare probe, $\Delta R$
might be of the same order of magnitude as the other sources, or even dominant.
In that case the total $v_2$ for $R$ should be significantly reduced or even
negative.

\begin{figure}[ht]
\centerline{
\includegraphics[width=3.0in,height=3.6in,angle=-90]{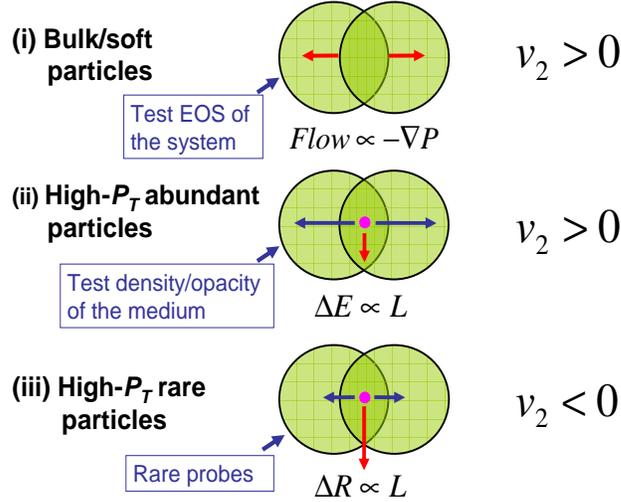}}
\caption{Three sources of particle asymmetries produced in heavy ion 
collisions. (i) $v_2 > 0$ for bulk particles generated by different pressure 
gradients in and out of the reaction plane. (ii) $v_2 > 0$ for quenched but
abundant high-$p_T$ partons generated by different length of propagation in 
and out of the reaction plane. (iii) $v_2 < 0$ for particles produced at 
high and intermediate $p_T$ from interactions of jets with the medium. The
asymmetry is also generated by different path lengths of propagating jets in 
and out of the reaction plane. All measured high-$p_T$ particles have
contributions from both (ii) and (iii) with the relative weight depending
on the initial abundance in jets and the chemical composition of the
medium.}
\label{fig:1}
\end{figure}

In Fig.\ \ref{fig:1} we show the three known sources of particle asymmetry
and the sign of $v_2$ that they imply. (i) $v_2$ can be generated by different
pressure gradients in and out of the reaction plane, leading to more
flow in the plane (i.e. in the ``thinner'' direction of the fireball)
implying $v_2 > 0$. This is elliptic flow in the strict sense of the word.
It is the dominant source of $v_2$ at low $p_T$.
(ii) High-$p_T$ partons lose more energy out of the plane (i.e.\ the
``thicker'' side of the fireball), so that jet quenching is less pronounced
in the plane, leading to $v_2 > 0$. This is the dominant source of $v_2$
for most high $p_T$-probes. (iii) Rare probes at high and intermediate $p_T$
are produced by interactions of jets with the medium. The longer the path
length in the medium, the more of these particles are produced, leading
to larger emission in the thicker direction of the fireball with $v_2 < 0$.
This has been called optical $v_2$ in \cite{Turbide:2005bz}.

Most theoretical calculations for the total direct photon $v_2$ predict
values which are close to zero or slightly negative at intermediate $p_T$.
PHENIX has presented first measurements which are compatible with zero
\cite{phenix:05v2} and rule out large negative elliptic flow of all
direct photons sources combined. More precise measurements in the future
will hopefully be able to provide tighter constraints.

In \cite{weiliu} we showed that strange quarks at RHIC energies could be an 
ideal probe in the sense discussed in this section, since $s$ and $\bar s$ 
quarks are rare as as leading particles of jets, but almost chemically 
equilibrated in the bulk medium formed in the collisions. This should result
in an increased yield of kaons an $\Lambda$ hyperons at high $p_T$ when
compared to scaled $p+p$ collisions. We conclude immediately that this
additional source of strange hadron exhibits negative $v_2$.

\section{$\mathbf{v_2}$ of Strange Quarks and Kaons}

In this section we make predictions for the azimuthal asymmetry of strange
quarks and kaons adding all known sources for these particles at intermediate
and large $p_T$. We use the method introduced in Ref.\ \cite{weiliu} to study
the propagation of jet partons ($u$, $d$, $s$, and $g$) in the QGP fireball. 
Energy loss and conversions are implemented through Fokker-Planck 
and rate equations. We use elastic perturbative cross sections between partons
scaled with a $K$-factor. We refer the reader to \cite{weiliu} for details.
After a jet parton has left the fireball we fragment into hadrons via AKK fragmentation function \cite{akk}, and
calculate the elliptic flow of the final state particles.

\begin{figure}[ht]
\includegraphics[width=3.0in,height=3.0in,angle=-90]{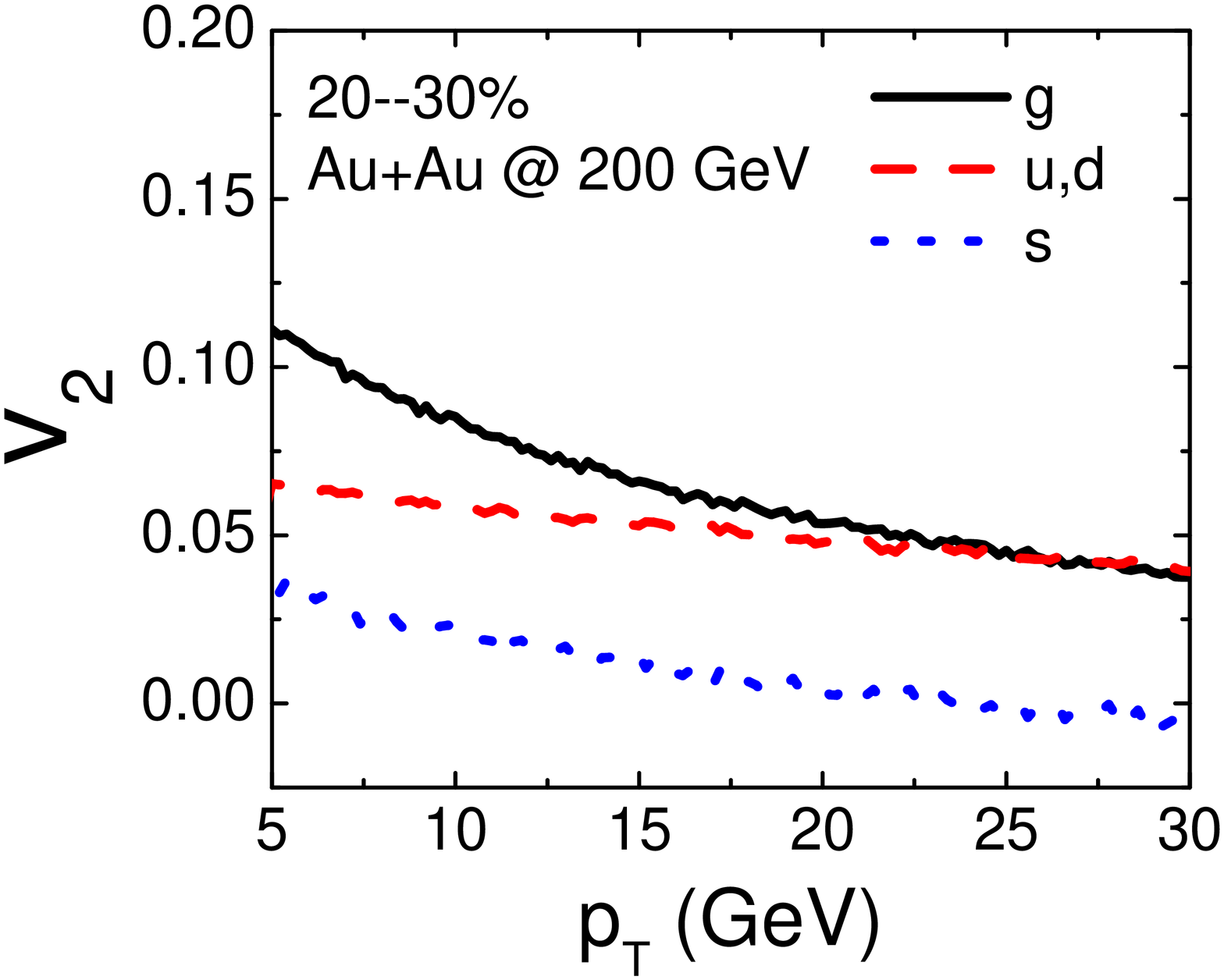}
\hspace{-0.2cm}
\includegraphics[width=3.0in,height=3.0in,angle=-90]{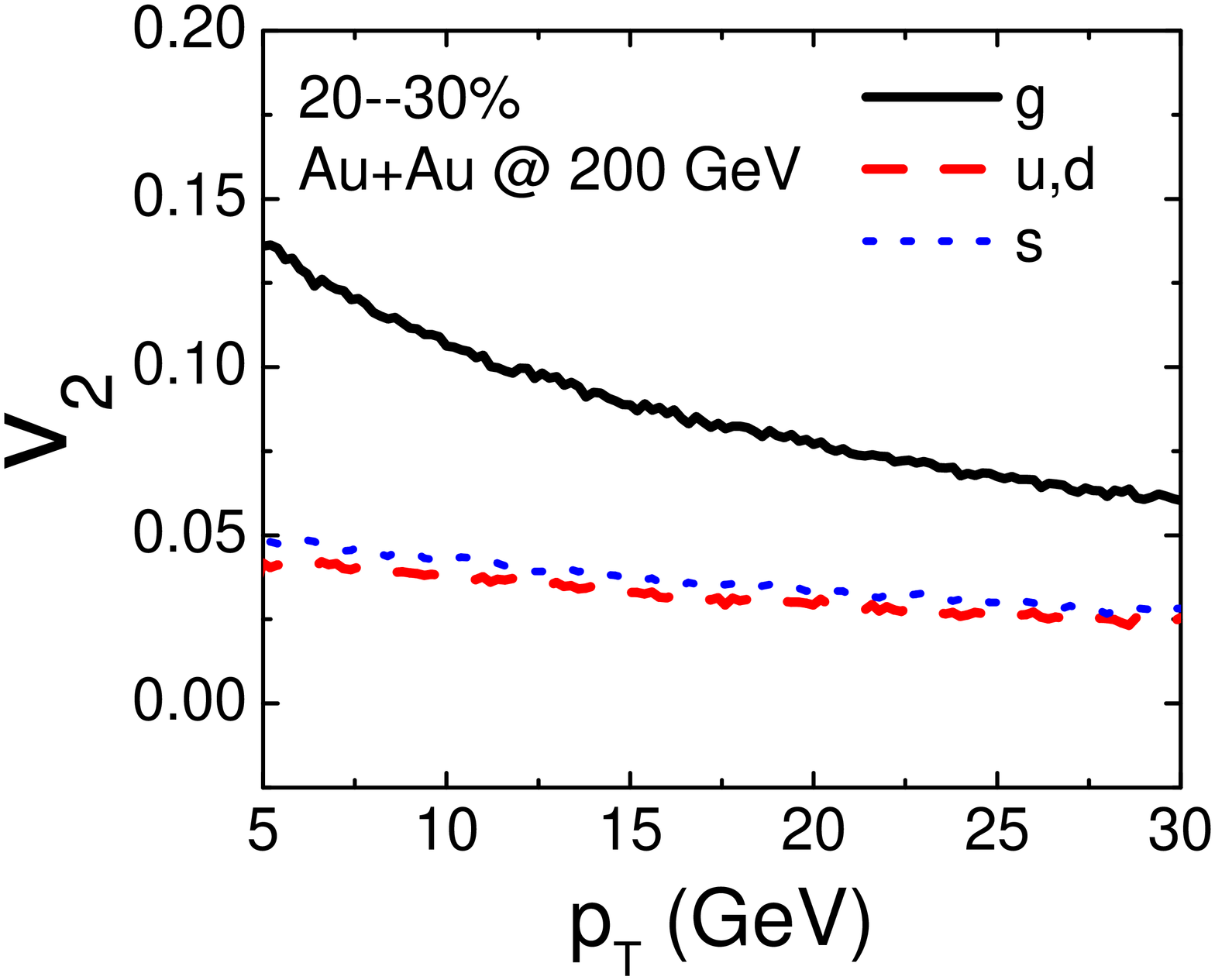}
\caption{Elliptic flow $v_2$ of partons with jet conversions
(left panel) and without jet conversions (right panel) as functions of
transverse momentum $p_T$ in Au+Au collisions at $\sqrt{s_{NN}}$ = 200 GeV.}
\label{fig:2}
\end{figure}

In Fig.\ \ref{fig:2}, we show the elliptic flow of light quarks ($u$, $d$),
strange quarks $s$ and gluons $g$ with conversions ($K=4$ for jet conversion
cross sections, left panel) and without conversions ($K=0$ for conversion
cross sections, right panel) in Au+Au collisions at $\sqrt{s_{NN}}=200$
GeV for the $20-30\%$ centrality bin. The elliptic flow of all partons
decreases with increasing transverse momentum $p_T$ consistent with the 
behavior of the drag coefficients discussed in \cite{weiliu}.
If the flavor of quark and gluon jets is kept fixed ($K=0$), gluons exhibit
larger elliptic flow than quarks. Jet conversions drive the $v_2$ of light
quarks and gluons toward each other as expected for particles that can easily
convert into each other. In complete detailed balance the curves should be the
same. On the other hand, $s$ quarks have the same $v_2$ as light quarks
with conversions switched off. When conversions are switched on, strange quark
$v_2$ drops significantly due to the large number of extra strange quarks
produced throughout the thicker part of the fireball.

\begin{figure}[ht]
\includegraphics[width=3.0in,height=3.0in,angle=-90]{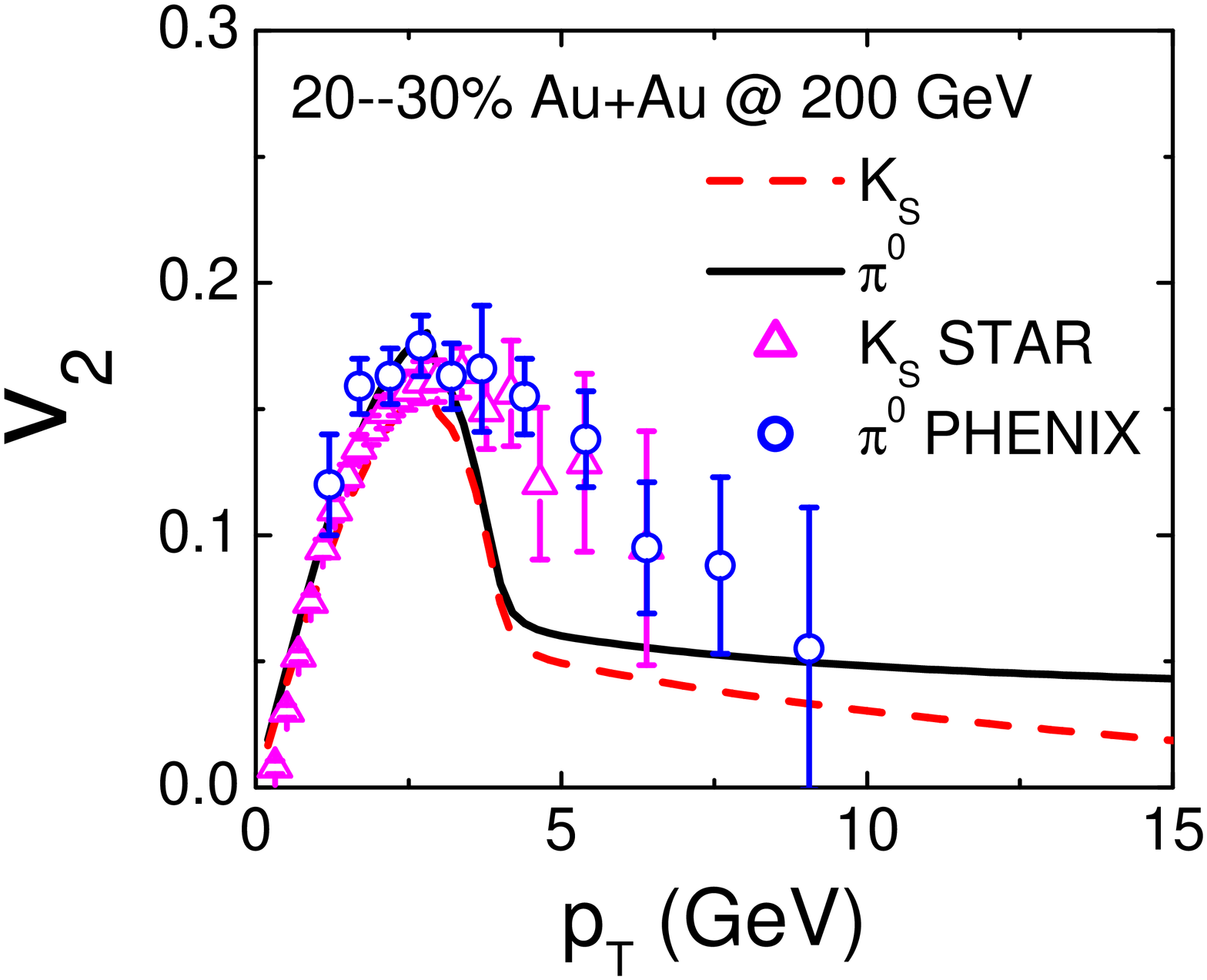}
\hspace{-0.2cm}
\includegraphics[width=3.0in,height=3.0in,angle=-90]{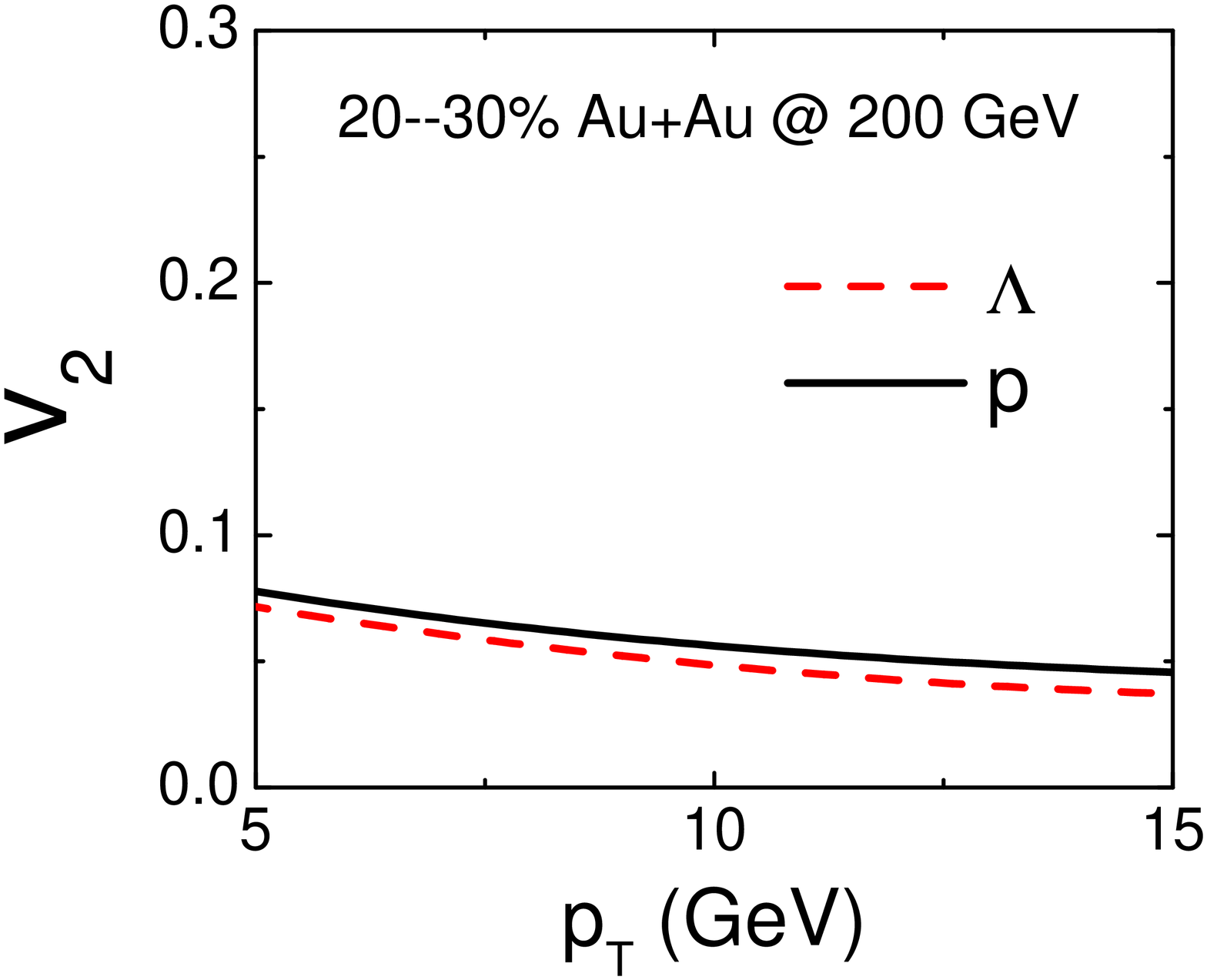}
\caption{Elliptic flow $v_2$ of $\pi^0$ and $K^0_S$ mesons (left panel) and of proton and $\Lambda$ baryons (right panel) including jet
conversions as functions of transverse momentum $p_T$ in Au+Au collisions
at $\sqrt{s_{NN}}$ = 200 GeV. Data is taken from \cite{david} and
\cite{abelev}.}
\label{fig:3}
\end{figure}

In the left panel of Fig.\ \ref{fig:3}, we show the resulting $v_2$ for neutral kaons and pions.
We have added hadrons from recombination of quarks in order to take into
account hadrons from sources other than jets which are increasingly important
for lower transverse momenta $p_T < 6$ GeV/$c$ \cite{greco,fries,Fries:2004ej,
hwa}.
We see that the $v_2$ for kaons is systematically below the $v_2$ for
pions due to jet conversions. At $p_T=15$ GeV/$c$ the difference can be as 
large as a factor of two. We also show experimental data which has very 
large error bars above 6 GeV/$c$ and does not extend beyond 10 GeV/$c$. 
For comparison, in the right panel of Fig.\ \ref{fig:3} we plot the 
resulting $v_2$ for protons and $\Lambda$ hyperons at high transverse 
momentum. We notice that the difference of elliptic flow between strange and
non-strange baryons is much smaller than that between strange and non-strange
mesons. This can be traced back to the dominance of gluon fragmentation 
for baryon production in the fragmentation functions. Thus, hadronization
dilutes the rather clear signal for strange quarks, but it appears that kaons
are nevertheless a good probe.

One point of caution here is the fact that jet quenching calculations (even
without resolving particle chemistry) have great difficulties to explain the
relatively large values of $v_2$ at large $p_T$ with realistic fireball
models, see e.g.\ \cite{Shuryak:2001me,Adams:2004wz}. A sign of this can 
also be seen in our calculation which underestimates the data between 4 and 
7 GeV/$c$. No final conclusion has been emerged on this phenomenon. However, 
we are confident that whatever the absolute value for the elliptic flow of 
pions, the $v_2$ of kaons should be suppressed relative to it.

We conclude with another thought. No calculation has been presented on the
$v_2$ of identified hadrons coming from changed multiplicities in jet cones
as advocated in \cite{Sapeta:2007ad}. Naively, we would expect that
this mechanism has less impact on the $v_2$ of strange hadrons. This might
lead the way to a possible distinction between both mechanisms.

\section{Summary}

We have discussed the fact that the azimuthal asymmetry of high-$p_T$ particles
created in interactions of jets with the surrounding medium in nuclear
collisions is negative. This should lead to a significant reduction in the
total $v_2$ observed for any rare high-$p_T$ probe, as already
demonstrated for the example of photons. Here, we showed for the first time
a prediction for the suppression of the $v_2$ of strange hadrons with
respect to non-strange hadrons at large $p_T$. The $v_2$ of kaons could
be as much as a factor of two smaller than that of pions, and could 
potentially be an unambiguous signal for jet conversions in the quark gluon 
plasma.

\ack            
We wish to thank Ralf Rapp and Che-Ming Ko for helpful discussions. This work 
was supported by RIKEN/BNL, DOE grant DE-AC02-98CH10886, and the Texas A\&M 
College of Science.

\end{document}